# Finding Near-Optimal Maximum Set of Disjoint $k$-Cliques in Real-World Social Networks

Wenqing Lin[†,◇], Xin Chen[‡,◇,*], Haoxuan Xie[§,*], Sibo Wang[‡], Siqiang Luo[§]

[†]*Tencent*, Shenzhen, China
[‡]*The Chinese University of Hong Kong*, Hong Kong SAR, China
[§]*Nanyang Technological University*, Singapore
edwlin@me.com, jerchenxin@gmail.com,
haoxuan001@e.ntu.edu.sg, swang@se.cuhk.edu.hk, siqiang.luo@ntu.edu.sg

*Abstract*—A $k$-clique is a dense graph, consisting of $k$ fully-connected nodes, that finds numerous applications, such as community detection and network analysis. In this paper, we study a new problem, that finds a maximum set of disjoint $k$-cliques in a given large real-world graph with a user-defined fixed number $k$, which can contribute to a good performance of teaming collaborative events in online games. However, this problem is NP-hard when $k \geq 3$, making it difficult to solve. To address that, we propose an efficient lightweight method that avoids significant overheads and achieves a $k$-approximation to the optimal, which is equipped with several optimization techniques, including the ordering method, degree estimation in the clique graph, and a lightweight implementation. Besides, to handle dynamic graphs that are widely seen in real-world social networks, we devise an efficient indexing method with careful swapping operations, leading to the efficient maintenance of a near-optimal result with frequent updates in the graph. In various experiments on several large graphs, our proposed approaches significantly outperform the competitors by up to 2 orders of magnitude in running time and 13.3% in the number of computed disjoint $k$-cliques, which demonstrates the superiority of the proposed approaches in terms of efficiency and effectiveness.

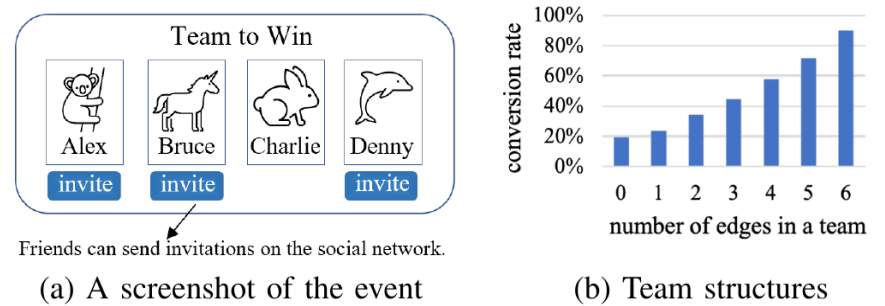

(a) A screenshot of the event  (b) Team structures

Fig. 1: A teaming event in a game

## I. INTRODUCTION

Given an undirected graph $G$, a $k$-clique in $G$ consists of exactly $k$ nodes in $G$, which are connected to each other. The $k$-clique is widely used in various applications, such as community detection [1]–[3] and social network analysis [4], [5]. Two $k$-cliques of $G$ are disjoint if they do not share any nodes in common. In this paper, we study a new problem, i.e., *the maximum set of disjoint $k$-cliques*, which finds a set of disjoint $k$-cliques, denoted by $\mathcal{S}$, in $G$ for a fixed $k \geq 3$ such that the number of $k$-cliques in $\mathcal{S}$ is maximum. Note that, the studied problem is NP-hard and cannot be addressed by the algorithms of maximum matching [6], as discussed in Section III.

Computing the maximum set of disjoint $k$-cliques can be found in various real-world applications, especially in social networks. For example, in a multiplayer online battle arena (MOBA) game of Tencent, which is one of the largest internet company in China, there are lots of teaming events to enhance the activities of players in the game using the social network, as shown in Figure 1(a). In particular, each player can join at most one team with up to 4 members, and is allowed to send invitations to friends in the same team to win the gaming rewards. For convenience, the game automatically generates all the teams and assigns each player to a team. The performance of the event is measured by the conversion rate, which is the ratio of players who win the rewards. As shown in Figure 1(b), players joining teams structured as $k$-cliques with 6 edges, i.e., all players in the team are friends to each other, exhibit the highest conversion rate, which outperforms the second-best teams with 5 edges by 25.6%. This can be attributed to the dense structure of $k$-cliques that facilitate more effective communication among all team members than non $k$-cliques. As a result, finding the maximum set of disjoint $k$-cliques could significantly enhance the performance of the teaming event. Note that, the maximum set of disjoint $k$-cliques could contain a large portion of nodes in the real-world graphs, e.g., 75% of the nodes in Dataset Orkut when $k = 4$. As for the uncontained nodes, the maximum set of disjoint dense-connected $k$ nodes can be found iteratively in the residual graph which removes the already contained nodes, until all nodes are settled. Besides, the number of $k$ could be sufficiently large, e.g., 5 in Honor of Kings[1] and 6 in QQ Speed Mobile[2]. More applications could be found, such as roommate allocation [7], where a room contains $k$ beds and a good arrangement is to make the roommates in a room form as a $k$-clique in the graph constructed based on their preferences, which is equivalent to finding the maximum set of disjoint $k$-cliques on the preference graph.

To address this problem, a straightforward approach takes three steps: (i) First, we list all $k$-cliques in the graph; (ii) Then, we construct a condensed graph, called *clique graph* (see Definition 2), by taking each $k$-clique as a condensed node and adding a condensed edge between two condensed nodes if the corresponding $k$-cliques are not disjoint; (iii) Finally, we adopt the algorithms for the maximum independent set (MIS) on the clique graph, which leads to the maximum set of disjoint $k$-cliques. However, this approach suffers from several deficiencies that render it impractical for handling large graphs, explained as follows. Firstly, the clique graph could be exponentially large and dense, e.g., the number of 3-cliques (resp. the condensed edges) in the Facebook dataset is at least 400 (resp. 100 thousand) times than the number of nodes (resp. edges) in the graph as shown in Table I. Secondly, computing the maximum independent set on large dense graphs is highly expensive [8]. In the experiments, we find that this approach can merely handle the graphs with only thousands of nodes.

Due to the hardness of computing the maximum set of disjoint $k$-cliques, we aim to find the near-optimal solution with a theoretical performance guarantee. Instead of constructing the costly clique graph in the previous approach, our approaches do not need to materialize all the $k$-cliques, which significantly reduces the overheads of space consumption and running time. In particular, we develop a basic framework by starting from a node $v$ in the graph $G$, and identifying a $k$-clique $c$ incident to $v$. Afterwards, we remove all the nodes in $c$ from $G$, as well as the corresponding edges, resulting in a residual graph $G'$. Note that, if there are not any $k$-cliques incident to $v$, we remove only $v$ from $G$. We continue to identify the new $k$-clique in $G'$, until the residual graph is empty. Therefore, we only need to maintain the $k$-cliques incident to the processed nodes and well prune the computation on the other nodes that have been in the chosen $k$-cliques, which largely narrows down the search space, rendering it possible to handle the large graphs widely existing in the real-world applications.

While the proposed framework is simple, there are several issues requiring non-trivial techniques to improve its performance in terms of both efficiency and effectiveness. Recall that we process the nodes in $G$ sequentially, which poses a node ordering. As a result, the node ordering can greatly affect the performance of the proposed framework. To explain, consider that we process nodes in descending order of their degree in $G$, i.e., we start from the node with the largest degree among the unprocessed ones. Since a node $v$ with a large degree might be associated with a large number of $k$-cliques, the process of node $v$ would be able to prune a large portion of the search space due to disjointness. Furthermore, it might also result in the case where a small-degree node is difficult to be included in a $k$-clique, which might be pruned by the large-degree nodes, leading to the number of disjoint $k$-cliques being far from the maximum one. On the other hand, if we process the nodes in ascending order of their degree in $G$, the $k$-clique $c$ computed on a small-degree node could include some large-degree nodes, making $c$ incident to a massive number of the other $k$-cliques. In other words, this ordering could face the same issue as the previous ordering.

To address these issues, we propose to consider the node degree in the clique graph constructed in the aforementioned method, instead of the original graph. As such, we are able to identify a better ordering of nodes to generate the set of disjoint $k$-cliques by taking into account the relations between $k$-cliques, which is also adopted in the algorithms for MIS (see Section III). However, computing the exact node degree in the clique graph is still costly, due to that it requires constructing the clique graph, whose size is too large to be processed. To alleviate this issue, we devise an approach to estimate the degree of each node in the clique graph efficiently and effectively with tight lower and upper bounds. Specifically, we first calculate the number of $k$-cliques incident to each node in $G$ by performing the $k$-clique listing algorithm in $G$ without storing all the $k$-cliques. As a result, we are able to derive the estimation of the node degree in the clique graph for each $k$-clique $c$ by exploiting the number of $k$-cliques incident to the nodes in $c$, as well as their neighbors in $G$, which also provides tight lower and upper bounds of the node degree in the clique graph. Based on that, we develop an efficient pruning strategy using the estimation of $k$-clique's degree in the clique graph, which enables us to look ahead at the nodes going to be processed, rendering it powerful in reducing the search space. Besides, we show that the proposed approach can achieve a $k$-approximation to the optimal solution, which guarantees its effectiveness. In the experiments, our proposed approach significantly outperforms the competitors in terms of both the quality of results and the required running time. For instance, in the Orkut dataset consisting of 3 million nodes and 117 million edges, compared to the competitors with $k=6$, our proposed approach generates 13.3% more disjoint $k$-cliques and achieves a speedup by one order of magnitude.

Nevertheless, real-world graphs often change frequently. For instance, the social network in the MOBA game of Tencent has a sufficiently large number of edge insertions or deletions in a day, caused by the construction or destruction of friendships between players, proportional to at least $1\%$ of all edges in the graph. As a result, how to maintain the high-quality maximal set of disjoint $k$-cliques for dynamic graphs would be important, due to the requirements for both accuracy and efficiency in the aforementioned applications. To explain, the deletion of existing edges makes the computed results inaccurate, i.e., some nodes in the graph are not connected after the deletion of edges. On the other hand, the insertion of new edges could affect the results by producing more disjoint $k$-cliques. To address this issue, a straightforward approach is to re-compute the maximal set of disjoint $k$-cliques on the updated graph. However, this approach is highly costly, making it impractical for real-world applications, due to that (i) computing the maximum disjoint $k$-clique set is expensive, and (ii) the update of the graph could be frequent in the applications, which would require timely responses for queries. Therefore, we devise an efficient updating method that builds an indexing structure for each $k$-clique $C$ in the result set $\mathcal{S}$, which is a subset of $k$-cliques incident to

$C$. As such, when dealing with an update affecting $C$, we can identify a replacement of $C$ from the index and update the index accordingly, which is significantly faster than re-computing from scratch. Our dynamic approach is shown to be both efficient and effective based on experimental results. For instance, in the Orkut dataset, when $k$ is 6, the average processing time for each update is just a few microseconds. Besides, the size of the updated $\mathcal{S}$ even increases slightly due to incurring more disjoint $k$-cliques.

**Contributions**. In summary, the contributions are as follows.

- We study the new problem of finding the maximum set of disjoint $k$-cliques in graphs, which has various applications in social networks. Besides, we show that this problem is NP-hard. (Section II)
- We devise a lightweight method that avoids the significant overheads and achieves a $k$-approximation to the optimal, which is equipped with several optimization techniques, including the ordering method, degree estimation in the clique graph, and a lightweight implementation. (Section IV)
- We extend the proposed approaches to handle the dynamic graphs, which is required by real-world applications. We develop an efficient indexing method with careful swapping operations that allows us to maintain a near-optimal result under sufficiently fast updates of edges. (Section V)
- We demonstrate in various experiments that our proposed approach is able to handle large graphs with millions of nodes, and performs much better than several competitors in terms of both efficiency and effectiveness. In particular, our proposed approach achieves up to 13.3% more $k$-cliques and consumes up to two orders of magnitude less running time than the competitors, making it suitable to handle large graphs. (Section VI)

## II. PRELIMINARIES

In this section, we introduce the basic concepts and the problem definition. Let $G = (V, E)$ be an undirected graph, where $V$ is the set of nodes with cardinality $n = |V|$, $E$ is the set of edges with cardinality $m = |E|$, and each edge connecting nodes $u$ and $v$ is denoted by $\langle u, v \rangle$. We first define the core concept, $k$-clique, as follows.

**Definition 1** ($k$-Clique). *A $k$-clique $C$ is a graph with exactly $k$ nodes, denoted by $C = (u_1, u_2, ..., u_k)$, which are incident to each other in the graph. That is, we have $\langle u, u' \rangle \in E$ for any $u \in C$ and $u' \in C$.*

Note that, $k$ is a user-defined fixed parameter and should be no less than 3. We say that two $k$-cliques $C_1$ and $C_2$ are *disjoint*, if there does not exist a node $v$ that exists in both $C_1$ and $C_2$. Based on that, we define the aforementioned *clique graph*, as follows.

**Definition 2** (Clique Graph). *Given all $k$-cliques in $G$, the clique graph of $G$, denoted by $G_C$, is an undirected graph, where (i) each $k$-clique of $G$ is a node in $G_C$, and (ii) there is an edge connecting any two nodes in $G_C$ if the corresponding $k$-cliques are not disjoint.*

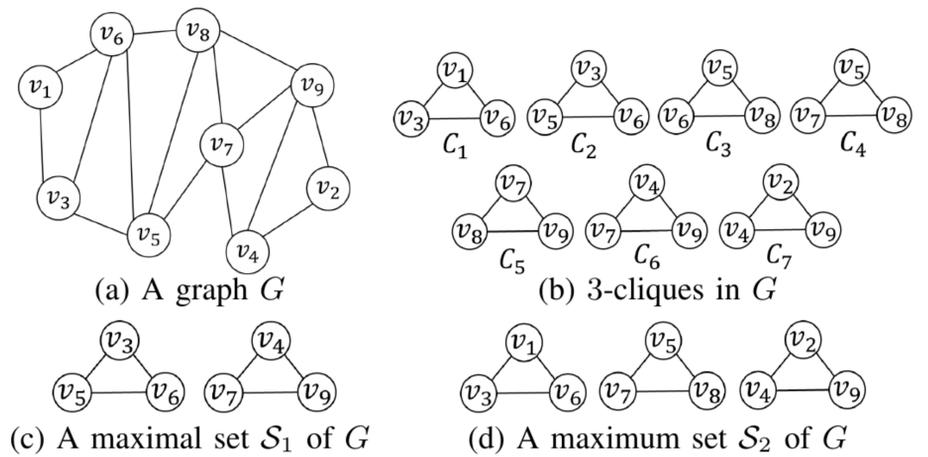

(a) A graph $G$    (b) 3-cliques in $G$

(c) A maximal set $\mathcal{S}_1$ of $G$    (d) A maximum set $\mathcal{S}_2$ of $G$

Fig. 2: A running example.

Therefore, we have the definition of *disjoint $k$-clique set*, stated in the following.

**Definition 3** (Disjoint $k$-Clique Set). *A disjoint $k$-clique set, denoted by $\mathcal{S}$, is a set of $k$-cliques such that each clique in $\mathcal{S}$ is disjoint with all other cliques in $\mathcal{S}$.*

The size of a disjoint $k$-clique set $\mathcal{S}$ is defined as the number of $k$-cliques in $\mathcal{S}$. We say that $\mathcal{S}$ is *maximal* if we can not add any other $k$-clique in $G$ into $\mathcal{S}$ without violating the disjoint constraint. Furthermore, we say that $\mathcal{S}$ is *maximum* if its size is the largest among all disjoint $k$-clique sets in $G$. Note that, the maximum disjoint $k$-clique set is a maximal one and not unique, i.e., there could be several different disjoint $k$-clique sets of the largest size.

**Example 1.** In Fig. 2(a) and 2(b), there is a graph $G$ with 9 nodes and 15 edges, which leads to seven 3-cliques in $G$, i.e., $(v_1, v_3, v_6)$, $(v_3, v_5, v_6)$, $(v_5, v_6, v_8)$, $(v_5, v_7, v_8)$, $(v_7, v_8, v_9)$, $(v_4, v_7, v_9)$, and $(v_2, v_4, v_9)$, denoted by $C_1, C_2, \cdots, C_7$ respectively. Besides, Fig. 2(c) and 2(d) show two disjoint 3-clique sets of $G$, denoted by $\mathcal{S}_1$ and $\mathcal{S}_2$ respectively. Note that, $\mathcal{S}_1$ and $\mathcal{S}_2$ are maximal due to that we cannot include any other disjoint 3-clique to these two sets, and $\mathcal{S}_2$ is maximum since the size of $\mathcal{S}_2$ is the largest. Furthermore, we can construct the clique graph $G_C$ of $G$, as shown in Fig. 3. Two nodes in $G_C$ have an edge only if the two corresponding cliques have common nodes. For example, $C_1$ and $C_2$ share the node $v_3$, resulting in an edge $\langle C_1, C_2 \rangle$ in $G_C$. □

Finding the maximum set of disjoint $k$-cliques is extremely costly, which can be explained in the following theorem.

**Theorem 1.** *The problem of finding the maximum set of disjoint $k$-cliques is NP-hard when $k \geq 3$ and is fixed.*

*Proof.* We construct the proof by reducing the exact cover by $k$-sets problem, denoted by $XkC$, which is NP-hard [9]–[12], to the maximum disjoint $k$-clique problem. Given a $k$-uniform hypergraph consisting of $n$ nodes, where each edge contains exactly $k$ nodes, $XkC$ finds a subset of disjoint hyperedges that cover all nodes. Suppose there is a $k$-uniform hypergraph $G_H = (V_H, E_H)$, we can build a new graph $G = (V, E)$ by regarding each hyperedge $\langle v_1, v_2, ..., v_k \rangle$ as a $k$-clique in $G$. In detail, for each hyperedge, we add its nodes to $V$ and add an edge $\langle u, v \rangle$ to $E$ between any two nodes $u$ and $v$ in this hyperedge. Building $G$ costs $O(|E_H| \cdot \binom{k}{2})$ time, which is in polynomial time when $k$ is fixed. As a consequence, if we can find a maximum set of disjoint $k$-cliques in $G$ that covers all nodes, it equals to a solution for the *XkC* problem, which completes the proof. □

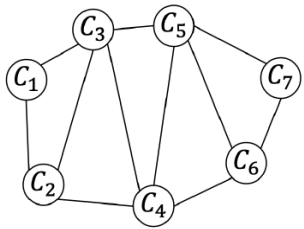

Fig. 3: The clique graph $G_C$ of $G$ in Fig. 2.

Therefore, it is highly difficult to compute the maximum set of disjoint $k$-cliques, not to mention that the graph could be sufficiently large. To approach this problem, we resort to the approximate solution that is able to generate a near-optimal maximum set of disjoint $k$-cliques with a theoretical performance guarantee on its size compared to the optimal.

## III. Related Work

In this section, we review the related problems, i.e., listing $k$-cliques, the maximum independent set, the exact cover by $k$-sets, maximum matching, diversified maximal cliques, and graph partitioning.

**Listing $k$-cliques**. A plethora of previous work have studied the algorithm design for efficiently listing $k$-cliques [13]–[18], which can be roughly summarized with a two-step framework: (i) First, given the graph $G = (V, E)$, we construct a directed acyclic graph (DAG) of $G$, denoted by $\vec{G} = (V, \vec{E})$, by exploiting a total ordering of nodes, such that $\vec{G}$ is the same as $G$ except that each edge in $\vec{E}$ has a direction that points from the node with a small ordering to the other node with a large ordering. The out-neighbor of $u$ in $\vec{G}$ is connected with the out-going edge of $u$ in $\vec{G}$. (ii) Then, for each node $u \in V$, we enumerate all $k$-cliques incident to $u$ by initializing a set $C$ consisting of $u$ and recursively adding a node $v \notin C$ into $C$ if $v$ is the out-neighbor of all nodes in $C$ and the size of $C$ is less than $k$. Note that, with the total ordering of nodes, each $k$-clique of $G$ is generated only once in the framework, which effectively avoids the redundant computation and leads to a time complexity $O(k \cdot m \cdot (d/2)^{k-2})$ [13] where $d$ is the maximum node degree of $G$. However, these algorithms for listing $k$-cliques cannot address the problem of finding the maximum set of disjoint $k$-cliques, due to that the number of $k$-cliques could be exponentially large and it requires the computation of maximum independent set on the listed $k$-cliques as aforementioned.

**Exact cover by k-sets**. The exact cover by $k$-sets problem is a well-known NP-hard problem [9]–[12]. Björklund [12] gives a algorithm which runs in $O^*((1+k/(k-1))^{n(k-1)/k})$ time and the polynomial space. Koivisto [10] proposes a simple clever dynamic programming over subsets which shows that the exact cover by $k$-sets can be solved in $O^*(2^{n(2k-2)/\sqrt{(2k-1)^2-2\ln 2}})$ time, while its space consumption is exponential. Björklund [9] develops a randomized polynomial space algorithm in $O^*(c_k^n)$ time, where $c_3 = 1.496$, $c_4 = 1.642$, $c_5 = 1.721$. However, this problem can not address our studied problem since it requires listing all $k$-cliques to construct the $k$-uniform hypergraph, which is tremendously expensive.

**Maximum matching in $k$-uniform hypergraphs**. As explained in Theorem 1, the problem of maximum disjoint $k$-clique set is related to the problem of maximum matching in $k$-uniform hypergraphs by regarding each $k$-clique as a hyperedge. Since the problem is NP-hard, a plethora of approximate algorithms [19]–[30] have been proposed. Most of these approaches adopt the framework by sequentially inspecting each hypergraph in an ordering such that the performance gain at each inspection is maximized, leading to a computation cost of polynomial time, as well as a theoretical performance guarantee, e.g., $(\frac{k+2}{3})$-approximation [26]. Nevertheless, it is infeasible to adopt the approaches for maximum matching in $k$-uniform hypergraphs to address the problem of maximum disjoint $k$-clique set, due to that it requires listing all $k$-cliques to construct the $k$-uniform hypergraph, which is tremendously expensive as discussed previously.

**Maximum matching in general undirected graphs**. When $k = 2$, finding the maximum set of disjoint $k$-cliques is equivalent to finding the maximum matching in general undirected graphs. To address this problem, there exist numerous works [6], [31]–[34] with a tight bound on the running time complexity. [6] proposed a method that runs in $O(n^4)$ by shrinking *Blossom* [6]. [32], [33], and [34] further improved the blossom-based method to $O(n^3)$, $O(n^{2.5})$, and $O(\sqrt{n} \cdot m)$, respectively. Instead of using the blossom, [31] introduced methods using *Augmenting Path* [31], which run in $O(n \cdot m)$. However, this problem is a special case of our studied problem, which is a more generalized case.

**Diversified maximal cliques**. The problem of diversified maximal cliques [35]–[37] aims to find at most a user-defined number of cliques to cover the most number of nodes, which is different from the studied problem of finding the maximum set of disjoint $k$-cliques in many aspects, explained as follows. First of all, the problem of diversified maximal cliques does not require that each clique has the same size with each other, e.g., $k$, which is a constraint in our studied problem. Secondly, one might argue that the output cliques of diversified maximal cliques can be easily divided into several small equal-sized cliques. However, this approach cannot lead to the maximum set of disjoint $k$-cliques, which generates the disjoint $k$-cliques as many as possible. Thirdly, our approach proposed the technique of clique degree estimation and ordering, which achieves the $k$-approximation and can be extended to handle the dynamic cases efficiently.

**Graph partitioning**. There are two categories of graph partitioning [38], [39], i.e., edge-cut and node-cut. Among them, the most related to our studied problem is the edge-cut graph partitioning, which divides the set of nodes into $t$ disjoint and

**Algorithm 1:** BASICFRAMEWORK($G$)

**Input:** An undirected graph $G = (V, E)$, a number $k$
**Output:** A maximal set $\mathcal{S}$ of disjoint $k$-cliques of $G$

1  $\mathcal{S} \leftarrow \emptyset$
2  Let $\eta$ be a total ordering on $V$
3  $\vec{G} = (V, \vec{E}) \leftarrow$ directed version of $G$, where $u \to v$ if $\eta(u) > \eta(v)$
4  $valid(v) \leftarrow true\ \forall v \in V$
5  **for each** node $u$ in ascending order of $\eta(u)$ **do**
6   **if** $valid(u) = true$ and $|N^+(u)| \geq k - 1$ **then**
7    **if** $FindOne(k-1, N^+(u), \{u\}) = true$ **then**
8     **for each** node $v$ in the newly found clique $C$ **do**
9      $valid(v) \leftarrow false$
10     Remove $v$ from $\vec{G}$
11    $\mathcal{S} \leftarrow \mathcal{S} \cup C$
12 **return** $\mathcal{S}$

13
14 **Procedure** $FindOne(l, V', C)$
15 **if** $l = 2$ **then**
16  Find an edge $\langle u, v \rangle$ of $\vec{G}$ and form a $k$-clique $C \cup \{u, v\}$
17  **return** true
18 **else**
19  **for each** node $u \in V'$ **do**
20   **if** $|N^+(u)| < l - 1$ **then**
21    **continue**
22   **if** $FindOne(l-1, N^+(u), C \cup \{u\}) = true$ **then**
23    **return** true
24 **return** false

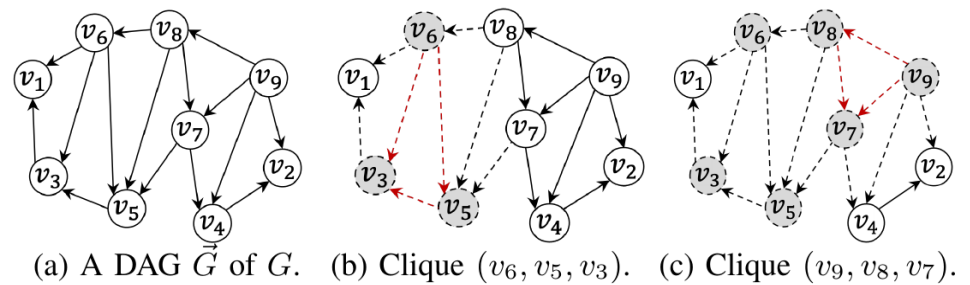

(a) A DAG $\vec{G}$ of $G$.  (b) Clique $(v_6, v_5, v_3)$.  (c) Clique $(v_9, v_8, v_7)$.

Fig. 4: A running example of Algorithm 1 on $G$ in Fig. 2.

even-sized subsets of nodes, where $t$ is a user-defined number, such that the number of cutting edges connecting different subsets is minimized. However, the graph structure induced on each subset produced by graph partitioning algorithm cannot be guaranteed to be dense, which is essential to the applications in real-world social networks that require the graph structure inside each subset to be as dense as possible to facilitate effective and efficient communications among nodes in the subset, as explained in Fig. 1.

## IV. PROPOSED ALGORITHMS

In this section, we first present a basic framework that avoids the overheads in the clique graph construction (Section IV-A). After that, we illustrate the optimization techniques based on the framework to generate a near-optimal result efficiently. To generate a near-optimal result, we propose the ordering method (Section IV-A) and degree estimation in the clique graph (Section IV-B). Further, to reduce the significant overheads in the computation and memory consumption as well as preserve a near-optimal result, we develop a lightweight implementation (Section IV-C).

### A. Basic Framework

As discussed in Section I, given a graph $G = (V, E)$, a straightforward approach is to (i) construct the clique graph by listing all $k$-cliques of $G$ and then (ii) apply the algorithms for MIS on the clique graph to generate the maximum disjoint $k$-clique set of $G$. However, this approach is highly costly, rendering it impossible to handle sufficiently large graphs. To address this issue, we propose a framework that does not need to construct the expensive clique graph, which significantly reduces the overheads. In particular, the framework begins with an empty set $\mathcal{S}$. Then, it works in several iterations. In each iteration, we add a $k$-clique $C$ in $G$ to $\mathcal{S}$ if $C$ is disjoint with the $k$-cliques in $\mathcal{S}$, and remove all nodes in $C$ from $G$. The iteration terminates when the residual graph of $G$ is empty or has no more $k$-clique that can be added into $\mathcal{S}$. Finally, we obtain $\mathcal{S}$ as a maximal set of disjoint $k$-cliques of $G$.

Given a graph $G = (V, E)$, Algorithm 1 presents the pseudocode of the basic framework to compute a maximal set of disjoint $k$-cliques of $G$. It first converts $G$ into a DAG (Line 3), denoted by $\vec{G} = (V, \vec{E})$, using a total node ordering $\eta$, which is able to effectively avoid redundant computation, as explained in Section III. For each node $u \in V$, denote $N^+(u)$ as the set of out-neighbors of $u$ in $\vec{G}$. In other words, the ordering of nodes $v$ in $N^+(u)$ is smaller than the one of $u$, denoted by $\eta(u) > \eta(v)$. Besides, every node $u$ maintains a value $valid(u)$ (Line 4), which records whether this node has been included in some clique of $\mathcal{S}$. Then, for each node $u$ in ascending order, it calls the procedure $FindOne$ to find a $(k-1)$-clique among the nodes in $N^+(u)$ that is first encountered (Lines 5-11). Then, the first encountered $(k-1)$-clique and node $u$ form a $k$-clique. Given a parameter $l$, $FindOne$ returns immediately once it finds a $l$-clique (Lines 17 and 23). After $FindOne$ returns a clique $C$, it directly adds $C$ into $\mathcal{S}$ and updates the graph $\vec{G}$ by removing nodes of this clique from $\vec{G}$ (Lines 8-11).

**Example 2.** Fig. 4(a) shows a DAG $\vec{G}$ of $G$ in Fig. 2 with the node ordering as $\eta(v_i) < \eta(v_j)$ for any $1 \leq i < j \leq 9$. Algorithm 1 inspects each nodes in $\vec{G}$ in the node ordering. Consider $k = 3$. Since only nodes $v_6, v_7, v_8$, and $v_9$ has at least two out-neighbors in $\vec{G}$, we only need to inspect these nodes. When processing the node $v_6$ whose out-neighbors are $v_1, v_3$, and $v_5$, we can recursively identify a 3-clique $C_1 = (v_6, v_5, v_3)$. Then, we remove $C_1$ from $\vec{G}$ by deleting all edges incident to nodes in $C_1$, resulting in the graph in Fig. 4(b). After that, only node $v_9$ has at least two out-neighbors in the residual graph of $\vec{G}$, which would lead to a 3-clique $C_2 = (v_9, v_8, v_7)$. By removing $C_2$ from $\vec{G}$ as shown in Fig. 4(c), there are no more nodes that should be inspected. The algorithm terminates and outputs the set of 3-

cliques containing $C_1$ and $C_2$, which is a maximal disjoint 3-clique set of $G$. □

It is worthy noting that Algorithm 1 does not list all $k$-cliques, due to that each selected $k$-clique results in the pruning of the search space. In other words, the selection of $k$-cliques would be essential to the performance of the proposed algorithm. Besides, the total ordering of nodes could affect the result of $\mathcal{S}$. To explain, suppose that we use the degree ordering, i.e., (i) a node with a larger degree has a larger total ordering, and (ii) ties are broken arbitrary if two nodes has the same degree. For each processed node $u$, we only need to consider the nodes in $N^+(u)$ for further computation, whose degrees are not larger than that of $u$. Thus, after adding a clique $C$ containing $u$ to $\mathcal{S}$ and removing nodes of $C$ from $\vec{G}$, the search space only decreases slightly, where the search space means the number of cliques in the remaining graph.

**Complexity analysis.** As Algorithm 1 inspects each node in $G$ and generates each $k$-clique at most once, its running time complexity should not be more than the one of listing $k$-cliques. That is, we have the time complexity of Algorithm 1 bounded by $O(k \cdot m \cdot (d/2)^{k-2})$ [13] where $d$ is the maximum node degree of $G$. On the other hand, due to that Algorithm 1 does not need to store all listed $k$-cliques, its space complexity is $O(m+n)$.

### B. $k$-Clique Ordering

As aforementioned, the selection of $k$-cliques to be added into $\mathcal{S}$ could significantly affect the size of $\mathcal{S}$. To address this issue, we propose an ordering of $k$-cliques of $G$ based on the node degree estimation of the clique graph of $G$, which would be able to generate a large number of disjoint $k$-cliques. Note that, the estimation does not need to construct the clique graph, which could avoid the overheads as mentioned previously. For ease of illustration, we first introduce the clique graph related concepts, and then explain the methodology for estimation with a theoretical guarantee.

We say that two $k$-cliques $C$ and $C'$ are *neighbors* in the clique graph $G_C$ if there exists a node $u$ such that $u$ appears in both $C$ and $C'$, i.e., $C$ and $C'$ are not disjoint. Based on that, we have the *degree* of a $k$-clique $C$, defined as follows.

**Definition 4** (Clique Degree). *Given a clique graph $G_C$ and a clique $C$ in $G_C$, the clique degree of $C$, denoted by $deg_{G_C}(C)$, equals the number of neighbors of $C$ in $G_C$.*

Recall that a straightforward approach is to compute the MIS on the clique graph, which iteratively adds the minimum-degree node to an initially empty solution while simultaneously removing the selected node and its neighbors from the graph until the graph is empty. Although this approach can lead to a near-optimal result by considering the degree of $k$-cliques in the clique graph, even such a simple heuristic method is not easily implemented due to the huge overheads of calculating the node degree in the clique graph, which requires the construction of the clique graph.

To address this issue, we devise a scoring method that effectively approximates the degree of each $k$-clique in the clique graph with a theoretical guarantee, as follows.

**Definition 5** (Node Score). *Given a node $u$ in $G$, the node score of $u$, denoted by $s_n(u)$, is the number of $k$-cliques containing $u$.*

Therefore, the score of a $k$-clique $c$ can be computed based on the scores of nodes in $c$, which is defined in the following.

**Definition 6** (Clique Score). *Given a clique $C$, the clique score of $C$, denoted by $s_c(C)$, equals the total score of nodes in $C$, i.e., $\sum_{u \in C} s_n(u)$.*

To further illustrate the above definitions, we provide the following examples.

**Example 3.** Consider the clique graph $G_C$ in Fig. 3, where $k=3$. The clique degree of each 3-clique can be computed in $G_C$, e.g., $deg_{G_C}(C_1) = 2$ due to that $C_1$ is incident to two 3-cliques $C_2$ and $C_3$ in $G_C$. Furthermore, we have the node score of $v_6$ as $s_n(v_6) = 3$, since there are three 3-cliques containing $v_6$, which are $(v_1, v_3, v_6)$, $(v_3, v_5, v_6)$, and $(v_5, v_6, v_8)$. Similarly, we have $s_n(v_5) = 3$ and $s_n(v_8) = 3$. Moreover, since the 3-clique $C_3$ contains three nodes $v_5$, $v_6$ and $v_8$, we have the clique score of $C_3$ as $s_c(C_3) = s_n(v_5) + s_n(v_6) + s_n(v_8) = 9$. □

As such, we can derive an upper bound and a lower bound of the clique degree for each $k$-clique $C$ in $G$ based on the clique score of $C$, which is stated as follows.

**Theorem 2.** *Given a $k$-clique $C$, the degree $deg_{G_C}(C)$ of $C$ in $G_C$ satisfies that $(s_c(C) - k)/(k-1) \le deg_{G_C}(C) \le s_c(C) - k$.*

*Proof.* Consider a $k$-clique $C$, which contains $k$ nodes, denoted by $v_1, v_2, ..., v_k$. For each node $u$ in $C$, let $\mathcal{S}(u)$ be the set of $k$-cliques containing $u$. Then, we have $C \in \mathcal{S}(u)$ and the size of $\mathcal{S}(u)$ is $s_n(u)$. As each $k$-clique $C'$ in $\mathcal{S}(u)$, excepting $C$, should be a neighbor of $C$ in the condensed graph due to that they have a common node $u$ with $C$, the contribution of $\mathcal{S}(u)$ to the degree of $C$ is at most $s_n(u) - 1$. Therefore, we have an upper bound of the degree of $C$ by taking into account the node scores of nodes in $C$, i.e., $deg_{G_C}(C) \le \sum_{u \in C}(s_n(u) - 1) = s_c(u) - k$, according to Definition 6. On the other hand, since each $k$-clique $C'$ in $\mathcal{S}(u)$ has $k-1$ nodes excepting $u$, meaning that $C'$ might be incident to $k-1$ other $k$-cliques which is not $C$, the contribution of $\mathcal{S}(u)$ to the degree of $C$ is at least $(s_n(u) - 1)/(k-1)$. As a result, we obtain a lower bound of $deg_{G_C}(C)$ as $\sum_{u \in C}(s_n(u) - 1)/(k-1) = (s_c(u) - k)/(k-1)$, which completes the proof. □

In other words, the degree of a $k$-clique $C$ is highly related to its clique score $s_c(C)$, which means that we can approximate the degree of $C$ in the clique graph by $s_c(C)$. Consequently, we can obtain the approximation degree of each $k$-clique in $G$ without explicitly constructing the condensed

**Algorithm 2:** COMPUTEWITHCLIQUESCORES($G$)

**Input:** An undirected graph $G = (V, E)$, a number $k$
**Output:** A set $\mathcal{S}$ of disjoint $k$-cliques of $G$

1   $\mathcal{S} \leftarrow \emptyset$
2   Store all $k$-cliques of $G$ in memory, calculate $s_c(C)$ for each clique $C$
3   **for each** *clique $C$ in the ascending order of $s_c(C)$* **do**
4       **if** *$C$ is disjoint with all $k$-cliques in $\mathcal{S}$* **then**
5           $\mathcal{S} \leftarrow \mathcal{S} \cup C$
6   **return** $\mathcal{S}$

graph, which would incur significant overheads in the computation and memory consumption. We further propose an algorithm that utilizes the clique score to generate a near-optimal result set $\mathcal{S}$.

Algorithm 2 presents the details of our proposed approach, which first stores all $k$-cliques in memory (Line 2) and then processes the $k$-cliques in the ascending order of their clique scores in several iterations (Lines 3-5). In each iteration, we add a $k$-clique $C$ into $\mathcal{S}$, if $c$ is disjoint with all $k$-cliques in $\mathcal{S}$. The algorithm terminates when all $k$-cliques are processed and returns $\mathcal{S}$ as the result.

**Analysis.** In the sequel, we first show that the proposed algorithm can achieve a $k$-approximation to the optimal. After that, we analyze the complexity of the proposed algorithm in terms of both space consumption and running time.

**Lemma 1.** *Given a clique graph $G_C$ and a node $C$ in $G_C$, if $C$ has at least $k+1$ neighbors in $G_C$, there exist two neighbors of $C$ in $G_C$ such that these two neighbors are connected with an edge in $G_C$.*

*Proof.* Suppose that there is a $k$-clique $C$. Any neighbor of $C$ must share at least one node of $C$. Thus, for its $k+1$ neighbors, there are at least two neighbors that share the same node of $C$, indicating that there should be an edge of $G_C$ connecting these two neighbors of $C$ in $G_C$. □

**Theorem 3.** *Any maximal $\mathcal{S}$ is a $k$-approximation solution.*

*Proof.* Suppose that we have a maximal disjoint $k$-clique set $\mathcal{S}$, by Lemma 1, when we remove a $k$-clique from $\mathcal{S}$, we can add up to $k$ of its neighbors to $\mathcal{S}$ without violating the disjoint constraint. Thus, the approximate ratio is $k$. □

Algorithm 2 can produce a near-optimal solution as every clique is chosen from a nearly global optimal view. However, this method is not efficient as it has to store all cliques in memory. Its time complexity is $O(k \cdot m \cdot (d/2)^{k-2} + \tau \cdot \log \tau)$, where $\tau$ is the number of all cliques. To explain, it needs to find a clique with the minimum clique score in each iteration, which requires an additional cost of $O(\tau \cdot \log \tau)$. Its space complexity is $O(m + n + \tau)$ as it needs to store all $k$-cliques.

### C. A Lightweight Implementation

Although Algorithm 2 does not need to build the condensed graph, it has the drawback that it requires computing and storing all the $k$-cliques in $G$ in memory to compute the score of each clique, which results in significant overheads in memory consumption. To explain, consider the Facebook dataset discussed in Section I, whose 3-cliques have a number at least 400 times than the number of its nodes. Therefore, storing all the $k$-cliques would explode the memory space, especially for large graphs. To alleviate this issue, we develop a lightweight implementation that (i) does not need to store all the $k$-cliques, and (ii) produces the same result as the one of Algorithm 2.

Algorithm 3 presents the proposed lightweight implementation. The main idea is first to find a $k$-clique with the local minimum clique score among the subgraphs induced on the set $N^+(u)$ of out-neighbors of each node $u$. Then, these locally identified $k$-cliques are collected, and a clique with the global minimum clique score among them is found in each iteration. In detail, it first calculates the node score for each node (Line 2) and then sets a total node ordering using the node score (Line 3). It is worth noting that calculating the node score can be directly performed during the enumeration of all $k$-cliques, requiring only a memory cost of $O(m+n)$, as we do not store any $k$-cliques in memory. After that, it converts the original graph $G$ into a DAG $\vec{G}$ using the total ordering (Line 4). To efficiently find the clique with the global minimum clique score, it uses a min-heap $MinHeap$ to maintain cliques with the local minimum clique score (Line 5). Then, it calls the procedure $HeapInit$ to initialize $MinHeap$ (Line 6), which calls the procedure $FindMin$, for each node $u$, to find the clique with the minimum clique score based on the set $N^+(u)$ of out-neighbors of $u$ and pushes this clique into $MinHeap$ (Lines 11-14). After initializing $MinHeap$, it calls the procedure $Calculation$ to produce $\mathcal{S}$ by finding a clique with the global minimum clique score in each iteration (Line 7). This procedure pops cliques from $MinHeap$ until $MinHeap$ is empty (Lines 32-39). During this step, when a clique is disjoint with all $k$-cliques in $\mathcal{S}$, then we add this clique to $\mathcal{S}$ and update $\vec{G}$ accordingly (Lines 34-35). Otherwise, if its internal node $u$ with the highest node ordering is still valid, it inspects a new clique with the minimum clique score based on $N^+(u)$ to $MinHeap$ (Lines 37-39).

We implement our score-driven pruning strategy in the procedure $FindMin$. The motivation behind this strategy is to prune certain branches when the sum of the node scores of the previous recursive nodes is no smaller than the minimum clique score that has been found, indicating that such branches can not produce a clique with a smaller clique score. In particular, $FindMin$ includes a parameter $S_{cur}$, which represents the sum of the node score of the previous recursive nodes. With this, we implement our score-driven pruning strategy (Lines 19-20 and Lines 27-28). Specifically, a branch is pruned when $S_{cur}$ plus the score of the currently visited node is no smaller than the minimum clique score that has been found.

**Analysis.** With converting $G$ into a DAG $\vec{G}$, the computed $k$-cliques induced on $N^+(u)$ for each node $u$ are disjoint, and their union exactly corresponds to the entire $k$-clique set.

**Algorithm 3:** LIGHTWEIGHT($G$)

    **Input:** An undirected graph $G = (V, E)$, a number $k$
    **Output:** A set $\mathcal{S}$ of disjoint $k$-cliques
1  $\mathcal{S} \leftarrow \emptyset$
2  Calculate $s_n(u)$ for each node $u \in V$ during the enumeration all $k$-cliques of $G$ (no need to store $k$-cliques in memory)
3  Let $\eta$ be a total ordering on $V$ such that if $\eta(u) < \eta(v)$, then $s_n(u) \leq s_n(v)$
4  $\vec{G} \leftarrow$ a directed version of $G$, where $u \to v$ if $\eta(u) > \eta(v)$
5  $MinHeap \leftarrow \emptyset$, $valid(v) \leftarrow false$ $\forall v \in V$
6  $HeapInit(MinHeap)$
7  $Calculation(MinHeap, \mathcal{S})$
8  **return** $\mathcal{S}$

9
10 **Procedure** $HeapInit(MinHeap)$
11  **for each** node $u \in V$ in parallel **do**
12    **if** $|N^+(u)| \geq k - 1$ **then**
13      $FindMin(l - 1, N^+(u), \{u\}, \emptyset, s_n(u))$
14      Push the output clique into $MinHeap$

15
16 **Procedure** $FindMin(l, V', C, C_{min}, S_{cur})$
17  **if** $l = 2$ **then**
18    **for each** node $u$ of $V'$ **do**
19      **if** $S_{cur} + s_n(u) \geq s_c(C_{min})$ **then**
20        **continue**
21      **for each** edge $\langle u, v \rangle$ of $\vec{G}$ **do**
22        $C_{new} \leftarrow C \cup \{u, v\}$
23        **if** $s_c(C_{new}) < s_c(C_{min})$ **then**
24          $C_{min} \leftarrow C_{new}$

25 **else**
26    **for each** node $u \in V(\vec{G})$ **do**
27      **if** $|N^+(u)| < l - 1$ or $S_{cur} + s_n(u) \geq s_c(C_{min})$ **then**
28        **continue**
29      $FindMin(l - 1, N^+(u), C \cup \{u\}, C_{min}, S_{cur} + s_n(u))$

30
31 **Procedure** $Calculation(MinHeap, \mathcal{S})$
32  **while** $MinHeap$ is not empty **do**
33    Pop a clique $C$ where node $u$ has the largest node ordering
34    **if** $C$ is disjoint with all $k$-cliques in $\mathcal{S}$ **then**
35      Repeat Lines 8-11 of Algorithm 1
36    **else**
37      **if** $valid(u) = true$ and $|N^+(u)| \geq k - 1$ **then**
38        $FindMin(l - 1, N^+(u), \{u\}, \emptyset, s_n(u))$
39        Push the output clique into $MinHeap$

Based on this observation, we can first find the clique with the local minimum clique score on $N^+(u)$ for each node $u$ in parallel and then identify the clique with the global minimum score among these found cliques. When an invalid clique from $N^+(u)$ is popped, we should update the clique with the local minimum clique score in $N^+(u)$, which incurs redundant computation. However, such redundant computation must happen with some nodes becoming invalid and removed from $N^+(u)$, which indicates the size of the computed directed graph gradually reduces. Furthermore, together with our score-driven pruning strategy, such redundant computation is limited. The time complexity of the algorithm is $O(n \cdot m \cdot (d/2)^{k-2})$. To explain, Algorithm 3 involves redundant $k$-clique listing computation where for each node $u$, it conducts computation in $N^+(u)$ at most $n/k$ times. The space complexity is $O(m+n)$. Additionally, Algorithm 3 can produce the same $\mathcal{S}$ as Algorithm 2, as analyzed in Theorem 4, while simultaneously reducing the time and space overheads.

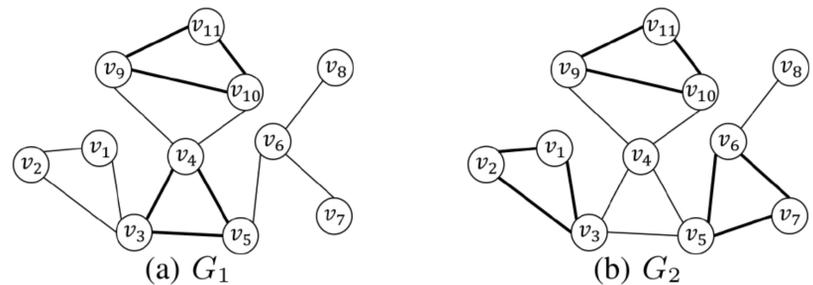

Fig. 5: Two graphs differing by only one edge with the bold edges in the corresponding maximum set of disjoint 3-cliques.

**Theorem 4.** *Given a fixed total node ordering and a fixed total ordering between cliques, Algorithm 3 and Algorithm 2 produce the same $\mathcal{S}$.*

*Proof.* A fixed total node ordering guarantees that $\vec{G}$ of the two algorithms are the same. A fixed total ordering between cliques guarantees a fixed ordering between any two cliques with the same clique score. As they always find the $k$-clique with the minimum clique score in each iteration, they produce the same $\mathcal{S}$. □

## V. HANDLING DYNAMIC GRAPHS

Previously, we assume that the graph is static, while most real-life graphs are dynamic with frequent updates by inserting new edges or deleting existing edges, as observed in Section I. Note that, the case of updates on the nodes can be treated equivalently as the updates on the edges incident to the corresponding nodes. Handling updates efficiently for our studied problem would be of great importance, due to that (i) some $k$-cliques incident to deleted edges could no longer exist in the updated graph, (ii) the result set $\mathcal{S}$ might no longer be the largest one, especially after inserting a sufficient number of edges, and (iii) the real-world applications would require a timely response for the updates.

To achieve that, we propose an efficient strategy by first inspecting a sufficiently small set $S$ of $k$-cliques that would be affected by the updates, and then swapping $k$-cliques in $\mathcal{S}$ with the ones in $S$ that would maximize the size of the result set $\mathcal{S}$. In order to make the inspection efficient, we develop an indexing approach that can easily retrieve the set $S$ according to the updates. In the following, we first present the algorithm designed for efficient swap operations in Section V-A, and then we introduce the indexing structures in Section V-B, after which we put all together by explaining the algorithms for

## Algorithm 4: TRYSWAP($q$)

**Input:** A directed graph $\vec{G}$, a number $k$, a FIFO queue $q$ of cliques in $\mathcal{S}$
**Output:** A set $\mathcal{S}$ of disjoint $k$-cliques

1 **while** $q$ is not empty **do**
2    $C \leftarrow q.pop()$
3    Locate all neighboring candidate cliques $\mathcal{C}(C)$ of $C$
4    Use Algorithm 2 to find disjoint $k$-cliques $S_{dis}$ among $\mathcal{C}(C)$
5    **if** $|S_{dis}| > 1$ **then**
6      Remove $C$ from $\mathcal{S}$ and add cliques in $S_{dis}$ to $\mathcal{S}$
7      Update the candidate cliques
8      Push any cliques $C'$ in $\mathcal{S}$ whose $\mathcal{C}(C')$ involves new candidate cliques to $q$
9 **return** $\mathcal{S}$

## Algorithm 5: CONSTRUCTION($\mathcal{S}$)

**Input:** A directed graph $\vec{G}$, a number $k$, a set $\mathcal{S}$ of disjoint $k$-cliques
**Output:** A set $\mathcal{C}$ of candidate $k$-cliques

1 **for each** clique $C$ in $\mathcal{S}$ in parallel **do**
2    Let $B$ be the nodes both in $C$ and $N_F(C)$
3    Find all $k$-cliques $C'$ except $C$ on $B$ and add $C'$ to $\mathcal{C}$
4 **return** $\mathcal{C}$

handling the updates based on the swapping operations and the indexing structures in Section V-C.

### A. Swap Operations

The main idea behind our swap operation is to remove a clique from $\mathcal{S}$ and add as many of its disjoint neighboring cliques as possible to $\mathcal{S}$ in order to increase the size of $\mathcal{S}$. To explain, given a graph $G = (V, E)$ and a set $\mathcal{S}$ of disjoint $k$-cliques of $G$, we first identify the nodes, called *free nodes*, that are not contained in any clique $C \in \mathcal{S}$. As such, the $k$-cliques, containing both free nodes and non-free nodes, are the interesting ones that could lead to increase of the size of $\mathcal{S}$ after the swap operations. We call such $k$-cliques as *candidate $k$-cliques*, and denote the set of all candidate $k$-cliques of $G$ by $\mathcal{C}$. The implication of a candidate $k$-clique $C'$ are twofold: (i) $C'$ should not contain only free nodes, otherwise $C'$ should be added to $\mathcal{S}$, which contradicts that $\mathcal{S}$ is a maximal set of disjoint $k$-cliques. (ii) On the other hand, $C'$ should not contain only non-free nodes, otherwise the swap operation of $C'$ cannot proceed due to that the non-free nodes are already contained in some cliques in $\mathcal{S}$. In light of that, the non-free nodes in a candidate $k$-cliques should be contained in the same $k$-clique in $\mathcal{S}$. Given a $k$-clique $C \in \mathcal{S}$, we denote $\mathcal{C}(C)$ as the set of all the candidate $k$-cliques $C' \in \mathcal{C}$ such that $C$ and $C'$ have at least one node in common, i.e., $C$ and $C'$ are not disjoint. As shown in Fig. 5, if adding a new edge $\langle v_5, v_7 \rangle$ to $G_1$ resulting in the graph $G_2$, the candidate 3-cliques of the clique $C_1 = (v_3, v_4, v_5)$ are $(v_1, v_2, v_3)$ and $(v_5, v_6, v_7)$, where the latter one is a new formed clique. Therefore, when a $k$-clique $C \in \mathcal{S}$ is updated which requires the swap operations, we only need to consider the candidate $k$-cliques in $\mathcal{C}(C)$.

Algorithm 4 depicts the proposed swap operation. It utilizes a FIFO queue to store cliques in $\mathcal{S}$ that are eligible for swapping. During each iteration, it pops a clique $C$ from the queue and attempts to find a set $S_{dis}$ of disjoint $k$-cliques using Algorithm 2 among its candidate cliques (Lines 2-4). If it finds a non-empty $S_{dis}$, $C$ is removed from $\mathcal{S}$ and the cliques in $S_{dis}$ are added to $\mathcal{S}$ (Lines 5-6), thereby increasing the size of $\mathcal{S}$. Subsequently, the candidate cliques are updated, and any clique $C'$ in $\mathcal{S}$ with $\mathcal{C}(C')$ involving new candidate cliques is pushed to the queue for additional swapping (Lines 7-8).

**Analysis**. The swap operation is effective as it can expand $\mathcal{S}$ and lead to achieving a nearly local optimum after several iterations. The time complexity of Algorithm 4 is $O(k \cdot |\mathcal{S}| \cdot \binom{k \cdot d}{k} \cdot \log \binom{k \cdot d}{k})$. To explain, for each $k$-clique, the cost of listing its neighboring $k$-cliques is $O(\binom{k \cdot d}{k})$ and it costs $O(\binom{k \cdot d}{k} \cdot \log \binom{k \cdot d}{k})$ to find a disjoint $k$-clique set. Besides, the number of swaps is $O(k \cdot |\mathcal{S}|)$. The space complexity is $O(m + n + |\mathcal{S}| + |\mathcal{C}|)$.

### B. Maintaining Candidate $k$-Cliques

However, listing candidate $k$-cliques from scratch for each swap is expensive, that incurs significantly overheads in redundant computation. Thus, we maintain candidate $k$-cliques to reduce the redundant computation cost during each swap operation. We regard these candidate $k$-cliques as the *indexing structure* for swap operation. Algorithm 5 illustrates the proposed approach for finding all candidate $k$-cliques. Recall that the non-free nodes in a candidate $k$-clique $C'$ should come from the same $k$-clique $C$ in $\mathcal{S}$, i.e., at most one neighbor of $C'$ in the clique graph is contained in $\mathcal{S}$. As such, we need to find out all candidate cliques for each clique $C$ in $\mathcal{S}$ and collecting them to form a set $\mathcal{C}$ of candidate cliques. Given a clique $C$ in $\mathcal{S}$, we use $N_F(C)$ to denote the set of free nodes that are the neighbors of any node of $C$ in $G$. To identify candidate cliques of $C$, we first inspects the set $B$ of nodes consisting of both the nodes in $C$ and the corresponding free nodes in $N_F(C)$. Subsequently, we find all $k$-cliques on $B$ excepting $C$, and add them to $\mathcal{C}$ (Line 3). For example, in Fig. 5(a), consider the maximum disjoint 3-clique set of $G_1$, which is $\mathcal{S} = \{C_1 = (v_3, v_4, v_5), C_2 = (v_9, v_{10}, v_{11})\}$. As for $C_1$, we inspects the nodes in $C_1$ and $N_F(C_1)$, i.e., $v_3, v_4, v_5, v_1, v_2, v_6$, which leads to a candidate 3-clique of $C_1$, i.e., $(v_1, v_2, v_3)$. On the other hand, $C_2$ has no candidate cliques, as it has no neighboring free nodes.

**Analysis**. The size of candidate $k$-cliques is bounded by $|\mathcal{S}| \cdot \binom{k \cdot d}{k}$. The time complexity of Algorithm 5 is $O(|\mathcal{S}| \cdot \binom{k \cdot d}{k})$. The space complexity is $O(m + n + |\mathcal{S}| + |\mathcal{C}|)$. In practice, the quantity of candidate cliques is not expected to be large due to the strong constraint of the candidate clique, which requires that internal nodes are either free nodes or belong to the same clique in $\mathcal{S}$.

**Algorithm 6:** INSERTION($\langle u, v \rangle$)

**Input:** A directed graph $\vec{G}$, a number $k$, a new edge $\langle u, v \rangle$
**Output:** A set $\mathcal{S}$ of disjoint $k$-cliques

1 **if** *only one node $u$ is a free node* **then**
2     Find new candidate $k$-cliques containing $\langle u, v \rangle$
3     **if** *found* **then**
4        $C \leftarrow$ the clique in $\mathcal{S}$ containing $v$
5        Queue $q.push(C)$
6        $TrySwap(q)$
7 **else if** *$u$ and $v$ are both free nodes* **then**
8     **if** *free nodes can form a clique* **then**
9        Add this new clique to $\mathcal{S}$
10        Update the candidate cliques
11     **else**
12        Update the candidate cliques
13        Queue $q \leftarrow \emptyset$
14        Push all cliques $C'$ in $\mathcal{S}$ whose $\mathcal{C}(C')$ involves new candidate cliques to $q$
15        $TrySwap(q)$
16 **return** $\mathcal{S}$

**Algorithm 7:** DELETION($\langle u, v \rangle$)

**Input:** A directed graph $\vec{G}$, a number $k$, a deleted edge $\langle u, v \rangle$
**Output:** A set $\mathcal{S}$ of disjoint $k$-cliques

1 **if** *$u$ and $v$ are in a clique of $\mathcal{S}$* **then**
2     Let $C$ be the deleted $k$-clique in $\mathcal{S}$
3     Queue $q.push(C)$
4     $TrySwap(q)$
5 **else**
6     Delete candidate $k$-cliques containing $\langle u, v \rangle$.
7 **return** $\mathcal{S}$

### C. Handling Updates

We are now ready to propose our dynamic methods by applying the swap operation and the index. However, the challenges arise when dealing with edge insertions and deletions: determining (i) when to apply the swap operation and (ii) how to identify the set of cliques that have the potential to expand $\mathcal{S}$ as the input of our swap operation. Regarding the second challenge, after performing edge updates, we select the cliques in $\mathcal{S}$ that include new candidate cliques as input for the swap operation. Thus, the answer to the first challenge naturally emerges: the swap operation is applied only when we can identify cliques in $\mathcal{S}$ that involve new candidate cliques. Next, we formally introduce how our insertion and deletion methods handle the above challenges.

*1) Incremental Update:* An edge insertion may create new candidate cliques that can be used to expand $\mathcal{S}$ by our swap operation, or it may produce new cliques that can be directly added to $\mathcal{S}$. Algorithm 6 presents the details of our incremental method that deals with two cases. Suppose that the new edge is $\langle u, v \rangle$. The first case occurs when only one node $u$ is a free node, and the second case occurs when both $u$ and $v$ are free nodes. If neither $u$ nor $v$ are free nodes, nothing needs to be done. In the first case where only $u$ is a free node (Lines 1-6), we first find new candidate $k$-cliques containing $\langle u, v \rangle$ (Line 2). If new candidate cliques are found, we add the clique containing $v$ in $\mathcal{S}$ to the queue $q$ as the parameter of conducting $TrySwap$ to expand $\mathcal{S}$. When both $u$ and $v$ are free nodes (Lines 7-15), if free nodes can form a new clique among the neighbors incident to $u$, $v$ and their neighboring free nodes, we can directly add the new clique to $\mathcal{S}$ and update the corresponding candidate cliques. Notably, we do not conduct $TrySwap$ in this case as the other cliques in $\mathcal{S}$ will not involve new candidate cliques. When free nodes can not produce cliques (Lines 12-15), as this new edge may form new candidate cliques, we first find new candidate cliques. Then, we push any clique $C'$ in $\mathcal{S}$ whose $\mathcal{C}(C')$ involves new candidate cliques to $q$ and use $TrySwap$ to expand $\mathcal{S}$. For instance, in Fig. 5, consider adding an edge $\langle v_5, v_7 \rangle$ in $G_1$ leading to the graph $G_2$. Due to that $v_7$ is a free node, the insertion creates a new candidate 3-clique $(v_5, v_6, v_7)$ for $(v_3, v_4, v_5)$, which already has a candidate 3-clique in $G_1$, i.e., $(v_1, v_2, v_3)$. Then, $TrySwap$ removes $(v_3, v_4, v_5)$ from $\mathcal{S}$, and adds $(v_5, v_6, v_7)$ and $(v_1, v_2, v_3)$ to $\mathcal{S}$, resulting in the increase of the size of $\mathcal{S}$.

**Analysis.** The bottleneck of the insertion method is the $TrySwap$ procedure. Thus, the time complexity of Algorithm 6 is $O(k \cdot |\mathcal{S}| \cdot \binom{k \cdot d}{k} \cdot \log \binom{k \cdot d}{k})$. The space complexity is $O(m + n + |\mathcal{S}| + |\mathcal{C}|)$.

*2) Decremental Update:* An edge deletion $\langle u, v \rangle$ can lead to the splitting of a clique in $\mathcal{S}$ or candidate cliques containing this edge. Thus, there are two cases to consider. The first case occurs when both $u$ and $v$ are in a clique of $\mathcal{S}$, which results in the splitting of this clique in $\mathcal{S}$ and makes $\mathcal{S}$ not maximal. Thus, we can apply our swap framework in this split clique to expand $\mathcal{S}$. The second case occurs when $u$ and $v$ form candidate $k$-cliques. We only need to delete such candidate $k$-cliques. Algorithm 7 presents the details of the edge deletion method. In the case where $u$ and $v$ are in a clique $C$ of $\mathcal{S}$ (Lines 1-4), we push $C$ to the queue and perform $TrySwap$ to expand $\mathcal{S}$. Otherwise, if $u$ and $v$ are not in a clique of $\mathcal{S}$, we directly delete invalid candidate $k$-cliques containing $\langle u, v \rangle$ (Lines 5-6). For example, in Fig. 5, consider the deletion of edges $\langle 5, 7 \rangle$ from $G_2$ leading to the graph $G_1$. Note that, the affected clique $(v_5, v_6, v_7)$ does not have any candidate 3-cliques, due to that the incident clique $(v_3, v_4, v_5)$ has the node $v_3$, which is contained by the other 3-cliques in $\mathcal{S}$. As a result, we have $\mathcal{S} = \{(v_1, v_2, v_3), (v_9, v_{10}, v_{11})\}$, which is also a maximum disjoint 3-clique set in $G_1$.

**Analysis.** The bottleneck of the deletion method is also the $TrySwap$ procedure. Thus, the time complexity of Algorithm 7 is $O(k \cdot |\mathcal{S}| \cdot \binom{k \cdot d}{k} \cdot \log \binom{k \cdot d}{k})$. The space complexity is $O(m + n + |\mathcal{S}| + |\mathcal{C}|)$.

## VI. EXPERIMENTS

We conduct extensive experiments over 10 real-world graphs on a Linux machine with an Intel Xeon 2.10GHz

TABLE I: Statistics of datasets. ($K, M, B, T = 10^3, 10^6, 10^9, 10^{12}$)

| Name | Dataset | n | m | Number of $k$-cliques | | | |
|---|---|---|---|---|---|---|---|
| | | | | $k=3$ | $k=4$ | $k=5$ | $k=6$ |
| FTB | Football | 115 | 613 | 810 | 732 | 473 | 237 |
| HST | Hamsterster | 1.86K | 12.5K | 16.8K | 10K | 2.77K | 285 |
| FB | Facebook | 4K | 88K | 1.61M | 30M | 518M | 7.83B |
| FBP | FBPages | 28K | 206K | 393K | 837K | 2.19M | 6.1M |
| FBW | FBWosn | 63.7K | 817K | 3.5M | 13.3M | 46.5M | 145M |
| DS | Dogster | 260K | 2.15M | 5.17M | 28.5M | 131M | 475M |
| SK | Skitter | 1.7M | 11M | 28.8M | 149M | 1.18B | 9.76B |
| FL | Flickr | 1.7M | 15.6M | 548M | 26.7B | 1.07T | 33.6T |
| LJ | Livejournal | 5.2M | 48.7M | 311M | 11.4B | 589B | 28.2T |
| OR | Orkut | 3M | 117M | 628K | 3.22B | 15.8B | 75.2B |

CPU and 504GB memory. All algorithms are implemented in C++ and compiled using g++ with full optimization. By default, we use 64 threads. The runtime of any algorithm that exceeds 24 hours will be reported as out-of-time "OOT". We also report out-of-memory "OOM" if the memory cost of any algorithm exceeds the limit. In the experiments, we vary $k$ from 3 to 6, according to the setting of real-world applications, as explained in Section I.

A. Experimental Setup

**Datasets.** We test on 10 real graph datasets with different scales, as shown in Table I. All datasets are publicly available on KONECT [40] and Network Repository [41].

**Competitors.** In the experiment, we adopt 5 competitors: the exact solution, denoted by OPT, which is calculated by computing the exact MIS solution with [42] in the clique graph; the basic framework on $G$ (Algorithm 1), denoted by HG; the method with a $k$-clique ordering (Algorithm 2), denoted by GC; the lightweight ordering method without our proposed score-based pruning strategy (Algorithm 3), denoted by L; the lightweight ordering method with our proposed score-based pruning strategy (Algorithm 3), denoted by LP.

**Implementation details.** Notably, although Algorithm 3 and Algorithm 2 can produce the same $\mathcal{S}$ with a fixed total node ordering and a fixed total clique ordering (see Theorem 4), we do not maintain the fixed total clique ordering for efficiency considerations. Instead, we only guarantee that when a clique is added to $\mathcal{S}$, its clique score is currently the minimum. In other words, if two cliques have the same clique score, then we directly add the first countered one to $\mathcal{S}$. As a result, the quality of $\mathcal{S}$ between these two algorithms may differ slightly. Our implementation is available at https://github.com/jerchenxin/disjoint-k-clique.

B. Varying $k$

We first evaluate the performance of our proposed algorithms in running time, output quality and space consumption. Figure 6 shows the running time of each algorithm. To ensure fairness, the running time includes initialization and calculation time. First, OPT is inefficient. OPT runs OOT or OOM even on small datasets since computing the clique graph and the exact MIS is costly. Besides, we can see that the running time of OPT decreases as $k$ increases in FTB and HST since the size of FTB and HST is too small and its clique graph also gets smaller as $k$ increases. Second, HG is the most efficient, and its running time remains nearly the same with varying $k$. Third, GC is much slower than LP, and it runs OOM for five datasets when $k$ is large. For example, for FL, LJ, and OR, it runs OOM when $k$ is larger than 3. Fourth, compared with GC, L and LP are efficient and nearly one to two orders of magnitude faster than GC. Compared with LP, when $k$ is 3 and 4, HG is nearly 2X faster in most datasets. As $k$ increases, the running time of L and LP exhibits nearly exponential growth. Finally, compared with L, LP is more and more efficient when $k$ is larger than 3. For example, LP is nearly one order of magnitude faster in LJ when $k=6$.

Table II presents the quality of $\mathcal{S}$ of each algorithm. We use the size of $\mathcal{S}$ to represent the quality of $\mathcal{S}$, and the larger the size, the better the quality. Due to the same quality of $\mathcal{S}$ of L and LP, we only report the quality of $\mathcal{S}$ of LP. In this table, we report the exact size of $\mathcal{S}$ for OPT and HG, and the relative size for GC and LP, denoted by $\Delta$, which is the difference between the size of their $\mathcal{S}$ and that of HG. For example, when $k$ is 3 in FBP, the size of $\mathcal{S}$ of HG is 1602 and $\Delta$ of GC and LP is 163 and 164, respectively. Consequently, the size of $\mathcal{S}$ of GC and LP is 1765 and 1766, respectively. First, we can see that LP and OPT can produce nearly the same $\mathcal{S}$, which indicates that LP can produce a high-quality solution. Second, we can see that the size of $\mathcal{S}$ of GC and LP is nearly the same. The slight difference is due to the absence of a strict total node and clique ordering for efficiency considerations. Third, compared with HG, the size of $\mathcal{S}$ of LP is nearly up to 13.3% larger. For example, in OR, when $k$ is 6, the size of $\mathcal{S}$ of LP is 13.3% larger. In LJ, when $k$ is 6, the size of $\mathcal{S}$ LP is 11.7% larger. Thus, we conclude that LP can produce a near-optimal $\mathcal{S}$.

Table III summarizes the space consumption of each algorithm. Recall that the space complexities of HG and LP are both $O(m+n)$. As a result, the space consumption of HG and LP remains relatively small (up to 13.5GB) and increases only slightly as $k$ grows larger. However, compared to HG, the space consumption of LP is 1.2 to 15 times larger due to the additional data structures and the extra memory overhead caused by multi-threading. In contrast, the space consumption of GC is significantly higher as it needs to load all $k$-cliques in memory. For instance, for the SK dataset, it requires 152GB even when $k=5$. Moreover, as $k$ increases, the space consumption of GC grows exponentially, resulting in that GC runs out of memory on some datasets, such as FL, LJ and OR, when $k>3$. Furthermore, the space consumption of OPT is even greater than that of GC, as OPT requires storing the clique graph in memory, which is exceptionally large. For example, in the HST dataset with $k=5$, the space consumption of OPT is 26.4 times larger than that of GC.

C. Comparison with Exact Solution

In this experiment, we compare LP with OPT in 6 public datasets [40] and demonstrate the effectiveness of LP. Given that computing the exact solution is highly costly, we limit our evaluation to 6 smaller datasets of different sizes. As shown in Table IV, LP can produce a solution whose size is close to that of OPT. For these datasets, in most cases, LP can produce an optimal solution. Furthermore, it is noteworthy that even for such small datasets like Lizard, OPT runs out of time when $k = 3$, which also underscores the superior efficiency of LP. We also observe that the error ratio is at most 8%.

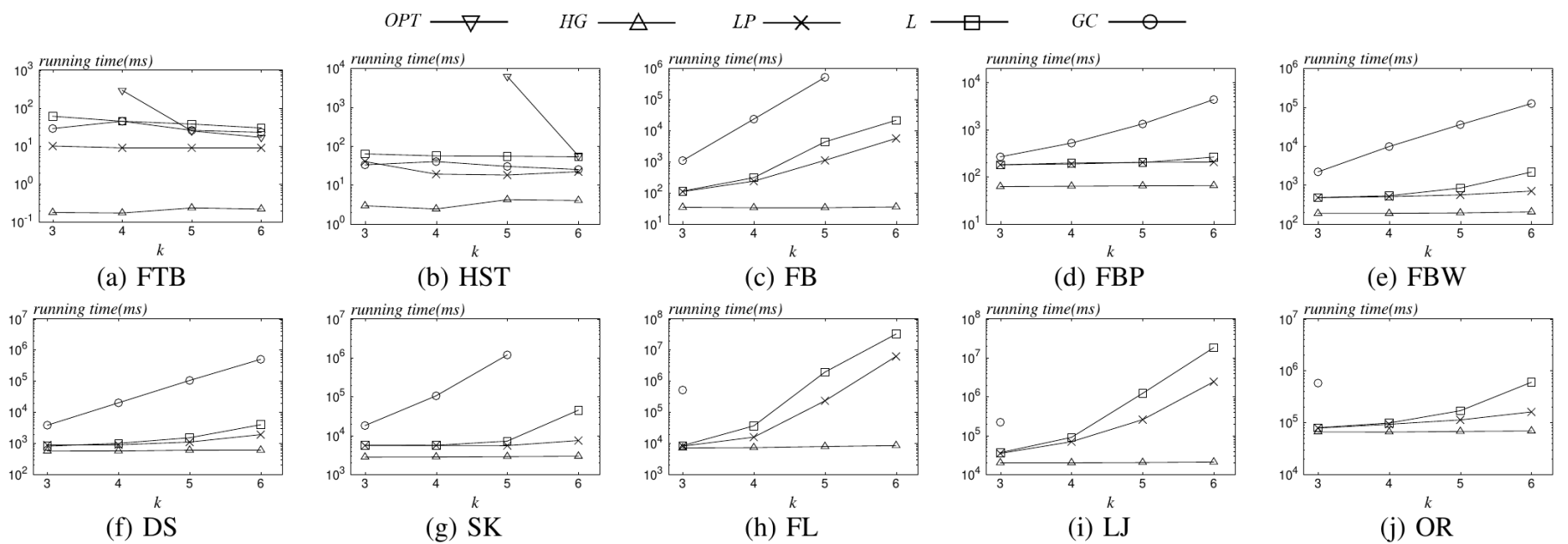

Fig. 6: Average running time in milliseconds with varying $k$.

TABLE II: Size of $\mathcal{S}$. (For GC and LP, $\Delta$ is the difference between the size of their $\mathcal{S}$ with that of HG)

| Name | k=3 | | | | k=4 | | | | k=5 | | | | k=6 | | | |
|---|---|---|---|---|---|---|---|---|---|---|---|---|---|---|---|---|
| | OPT | HG | GC ($\Delta$) | LP ($\Delta$) | OPT | HG | GC ($\Delta$) | LP ($\Delta$) | OPT | HG | GC ($\Delta$) | LP ($\Delta$) | OPT | HG | GC ($\Delta$) | LP ($\Delta$) |
| FTB | OOT | 32 | 4 | 4 | 25 | 24 | -1 | -1 | 16 | 16 | 0 | 0 | 11 | 11 | 0 | 0 |
| HST | OOT | 201 | 10 | 10 | OOT | 52 | 6 | 6 | 15 | 13 | 1 | 1 | 5 | 4 | 1 | 1 |
| FB | OOT | 1,195 | 40 | 40 | OOM | 784 | 48 | 48 | OOM | 561 | 37 | 37 | OOM | 413 | OOM | 31 |
| FBP | OOT | 5,732 | 357 | 348 | OOT | 2,888 | 254 | 249 | OOM | 1,602 | 163 | 164 | OOM | 967 | 88 | 106 |
| FBW | OOT | 13,114 | 1,112 | 1,121 | OOM | 7,041 | 661 | 660 | OOM | 4,146 | 443 | 447 | OOM | 2,606 | 307 | 309 |
| DS | OOM | 19,974 | 364 | 357 | OOM | 5,044 | 231 | 234 | OOM | 1,787 | 127 | 127 | OOM | 797 | 78 | 78 |
| SK | OOM | 132,009 | 4,458 | 4,495 | OOM | 31,775 | 2,054 | 2,058 | OOM | 10,320 | 985 | 987 | OOM | 4,354 | OOM | 480 |
| FL | OOM | 132,308 | 8,856 | 8,830 | OOM | 51,615 | OOM | 4,069 | OOM | 24,220 | OOM | 1,556 | OOM | 12,937 | OOM | 307 |
| LJ | OOM | 874,457 | 74,976 | 75,026 | OOM | 430,014 | OOM | 41,239 | OOM | 232,180 | OOM | 24,684 | OOM | 133,795 | OOM | 14,877 |
| OR | OOM | 861,315 | 54,590 | 54,556 | OOM | 513,758 | OOM | 49,093 | OOM | 323,078 | OOM | 38,041 | OOM | 212,440 | OOM | 28,186 |

TABLE III: Space consumption in megabytes (MB).

| Name | k=3 | | | | k=4 | | | | k=5 | | | | k=6 | | | |
|---|---|---|---|---|---|---|---|---|---|---|---|---|---|---|---|---|
| | OPT | HG | GC | LP | OPT | HG | GC | LP | OPT | HG | GC | LP | OPT | HG | GC | LP |
| FTB | OOT | 1.88 | 3.75 | 3.8 | 18 | 1.9 | 4.18 | 3.85 | 16 | 1.91 | 3.77 | 3.86 | 10 | 1.98 | 4 | 3.66 |
| HST | OOT | 3.55 | 4.8 | 4.34 | OOT | 3.52 | 4.48 | 4.36 | 115 | 3.61 | 4.36 | 4.32 | 21 | 3.55 | 4.3 | 4.39 |
| FB | OOT | 4.78 | 160 | 20.7 | OOM | 4.89 | 3880 | 21.5 | OOM | 4.73 | 67000 | 21.5 | OOM | 4.74 | OOM | 28.3 |
| FBP | OOT | 8.21 | 94.9 | 86.2 | OOT | 8.23 | 180 | 93.5 | OOM | 8.03 | 347 | 101 | OOM | 8.38 | 1040 | 107 |
| FBW | OOT | 19.2 | 423 | 221 | OOM | 18.9 | 1860 | 237 | OOM | 19.1 | 6150 | 251 | OOM | 19.1 | 23300 | 268 |
| DS | OOM | 49.1 | 1070 | 849 | OOM | 49.4 | 4170 | 911 | OOM | 50.2 | 17500 | 977 | OOM | 51 | 76000 | 1050 |
| SK | OOM | 267 | 5350 | 2920 | OOM | 266 | 22300 | 3350 | OOM | 271 | 156000 | 3870 | OOM | 277 | OOM | 4210 |
| FL | OOM | 322 | 49500 | 3920 | OOM | 321 | OOM | 4360 | OOM | 327 | OOM | 4740 | OOM | 333 | OOM | 5140 |
| LJ | OOM | 1030 | 35500 | 10200 | OOM | 1030 | OOM | 11500 | OOM | 1040 | OOM | 12700 | OOM | 1050 | OOM | 13800 |
| OR | OOM | 1800 | 60800 | 9310 | OOM | 1790 | OOM | 10100 | OOM | 1800 | OOM | 10800 | OOM | 1801 | OOM | 11500 |

### D. Performance on Synthetic Datasets

In this experiment, we vary the graph density to evaluate the scalability of HG, GC, and LP. We generate several random graphs under the Watts-Strogatz model [43] with the number of nodes equal to $n = 1M$, while the average node degree is varied from 8 to 64, leading to the number of edges varied from $4M$ to $32M$ respectively. Table V presents the running time, and Table VI presents the size of $\mathcal{S}$. We can first observe that both the size of $\mathcal{S}$ and the running time of each method increase as the graph becomes denser. Besides, the running time of HG remains nearly the same as $k$ increases. The running time of GC and LP depends on the number of $k$-cliques as collecting all $k$-cliques and the node score calculation usually consume most of the time. For example, when the average degree is 32, the running time of LP first increases and then decreases when $k \geq 4$.

### E. Evaluation on Dynamic Graphs

In this section, we evaluate the performance of our proposed dynamic algorithms in indexing time, index size, update performance, and the quality of $\mathcal{S}$ after updates.

First, we evaluate the indexing time and the index size of the proposed index. Table VII shows the indexing time and the index size, with the number of candidate cliques representing the index size. First, we can observe that indexing time increases with the increase of the index size. For example, in FBP, with the increase of $k$, the index size increases, leading to an increase in indexing time. Second, we can observe our index is efficient to build, and has a small size. For instance, in OR when $k$ is 6, it only has 1.92M candidate cliques while it has 75.2B 6-cliques. To explain, nodes in a candidate clique can be either free nodes or belong to the same clique in $\mathcal{S}$. Such a strict constraint limits the index size.

TABLE IV: Comparison with exact solution (ER is error ratio).

| Dataset | n | m | k=3 | | | k=4 | | | k=5 | | | k=6 | | |
|---|---|---|---|---|---|---|---|---|---|---|---|---|---|---|
| | | | LP | OPT | ER | LP | OPT | ER | LP | OPT | ER | LP | OPT | ER |
| Swallow | 17 | 53 | 4 | 4 | 0% | 2 | 2 | 0% | 0 | 0 | 0% | 0 | 0 | 0% |
| Tortoise | 35 | 104 | 6 | 6 | 0% | 2 | 2 | 0% | 1 | 1 | 0% | 1 | 1 | 0% |
| Lizard | 60 | 318 | 19 | OOT | - | 13 | 14 | 7.14% | 9 | 9 | 0% | 4 | 4 | 0% |
| Football | 115 | 613 | 36 | OOT | - | 23 | 25 | 8% | 16 | 16 | 0% | 11 | 11 | 0% |
| Voles | 181 | 515 | 48 | 49 | 2.04% | 30 | 30 | 0% | 18 | 18 | 0% | 13 | 13 | 0% |
| Hamsterster | 1.86K | 12.5K | 211 | OOT | - | 58 | OOT | - | 14 | 15 | 6.67% | 11 | 11 | 0% |

TABLE V: Running time (s) on synthetic datasets.

| Degree | k=3 | | | k=4 | | | k=5 | | | k=6 | | |
|---|---|---|---|---|---|---|---|---|---|---|---|---|
| | HG | GC | LP | HG | GC | LP | HG | GC | LP | HG | GC | LP |
| 8 | 1.2 | 3.28 | 3.98 | 1.22 | 2.29 | 3.24 | 1.19 | 1.9 | 2.77 | 1.14 | 1.49 | 1.96 |
| 16 | 2.2 | 9.36 | 6.98 | 2.41 | 9.96 | 6.14 | 2.3 | 6.26 | 5.34 | 2.4 | 4.1 | 4.3 |
| 32 | 4.15 | 35.5 | 14.3 | 4.29 | 116 | 15.4 | 4.45 | 146 | 15.3 | 4.64 | 123 | 13.6 |
| 64 | 9.85 | 194 | 32.2 | 8.78 | 1.66K | 47.3 | 8.86 | 4.83K | 82.4 | 9.14 | OOM | 133 |

TABLE VI: Size of $\mathcal{S}$ on synthetic datasets.

| Degree | k=3 | | | k=4 | | | k=5 | | | k=6 | | |
|---|---|---|---|---|---|---|---|---|---|---|---|---|
| | HG | GC ($\Delta$) | LP ($\Delta$) | HG | GC ($\Delta$) | LP ($\Delta$) | HG | GC ($\Delta$) | LP ($\Delta$) | HG | GC ($\Delta$) | LP ($\Delta$) |
| 8 | 275,636 | 19,394 | 19,164 | 160,284 | 14,764 | 14,693 | 57,106 | 717 | 719 | 0 | 0 | 0 |
| 16 | 303,408 | 7,277 | 7,291 | 206,693 | 17,764 | 17,695 | 138,749 | 24,804 | 24,783 | 88,509 | 12,026 | 12,050 |
| 32 | 317,417 | 3,509 | 3,444 | 228,178 | 7,460 | 7,446 | 171,736 | 12,707 | 12,710 | 130,654 | 18,749 | 18,735 |
| 64 | 324,964 | 1,629 | 1,683 | 238,631 | 3,278 | 3,273 | 185,736 | 5,421 | 5,405 | 149,210 | OOM | 8,131 |

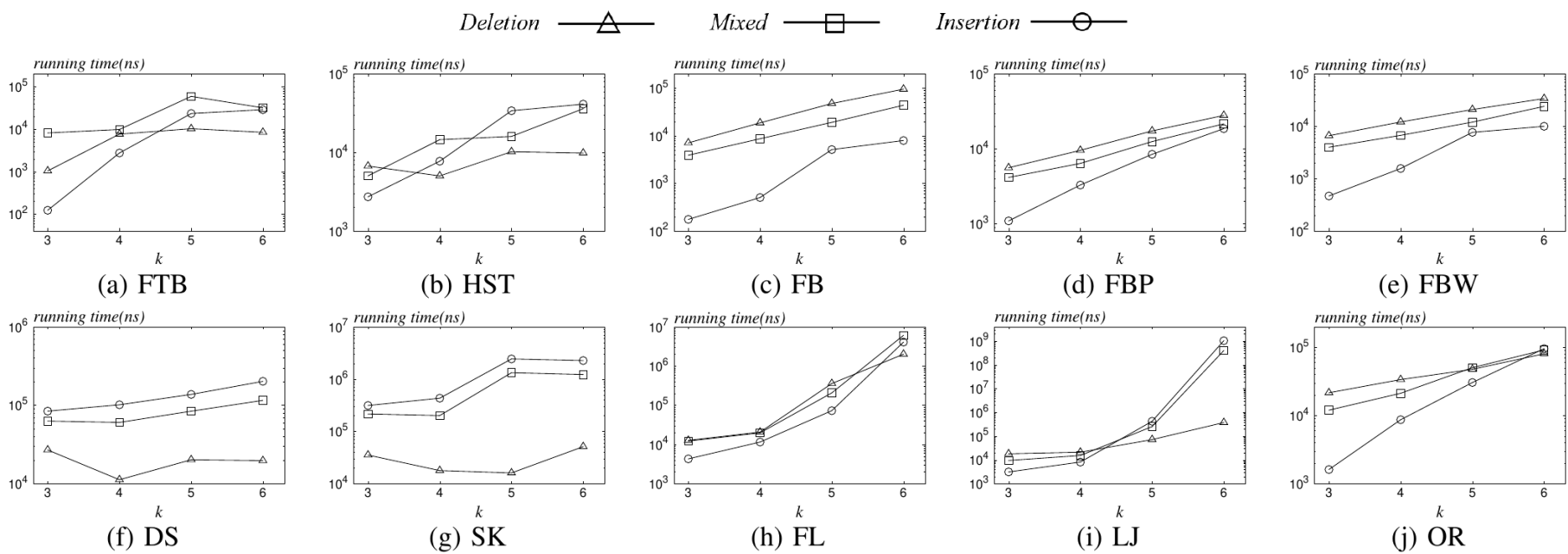

Fig. 7: Average update time in nanosecond seconds with varying $k$ in deletion, insertion, and the mixed workload.

Next, we evaluate the update performance under three workloads: 10K edge deletions, 10K edge insertions, and mixed workloads with 20K updates. For FTB and HST, we use 10 and 1K edge deletions and insertions, and mixed workloads with 20 and 2K updates, respectively, since the graph is small. For the first and second workloads, we first randomly and uniformly select 10K edges. Then, we delete these edges to report the performance for edge deletions and add them back to report the performance for edge insertions. For the mixed workloads, we generate 20K updates including 10K insertions and 10K deletions. All edges are also selected randomly and uniformly. For 10K insertions, we will first delete them from $G$ to form a new graph $G'$. Then, we evaluate the performance in the mixed workload by applying 20K updates in $G'$. For each workload, we report the average running time in Figure 7 and the size of $\mathcal{S}$ after updates in Table VIII.

TABLE VII: Indexing time and index size.

| Dataset | Indexing Time (ms) | | | | Index Size | | | |
|---|---|---|---|---|---|---|---|---|
| | 3 | 4 | 5 | 6 | 3 | 4 | 5 | 6 |
| FTB | 7.1 | 11.1 | 11.1 | 11.3 | 86 | 149 | 419 | 226 |
| HST | 9.91 | 17.7 | 11.6 | 15.5 | 1.01K | 327 | 274 | 20 |
| FB | 10.5 | 9.57 | 19.6 | 43.4 | 1.45K | 3.03K | 3.87K | 16.8K |
| FBP | 37.3 | 41.7 | 44.9 | 61.7 | 9.61K | 10.9K | 16.3K | 25.7K |
| FBW | 58.2 | 63.7 | 73.1 | 88.6 | 5.98K | 9.41K | 16.6K | 30.2K |
| DS | 307 | 207 | 192 | 190 | 109K | 15.3K | 5.18K | 3.36K |
| SK | 2.73K | 1.83K | 1.71K | 2.06K | 1.51M | 453K | 188K | 132K |
| FL | 1.46K | 1.52K | 1.84K | 2.66K | 173K | 103K | 274K | 419K |
| LJ | 5.59K | 5.8K | 7.16K | 14.6K | 983K | 1.13M | 3M | 4.73M |
| OR | 4.97K | 4.95K | 5.19K | 7.36K | 278K | 612K | 1.07M | 1.92M |

TABLE VIII: Quality of $\mathcal{S}$ after updates. ($\Delta$ is the difference of the size of $\mathcal{S}$ compared with that of building from scratch)

| Dataset | After Deletion ($\Delta$) | | | | After Insertion ($\Delta$) | | | | After Mixed Updates ($\Delta$) | | | |
|---|---|---|---|---|---|---|---|---|---|---|---|---|
| | 3 | 4 | 5 | 6 | 3 | 4 | 5 | 6 | 3 | 4 | 5 | 6 |
| FTB | 0 | 0 | 0 | -1 | 0 | 0 | 0 | 0 | 0 | 0 | 0 | 0 |
| HST | 0 | 3 | -1 | 0 | 0 | 2 | 1 | 0 | -1 | -1 | 0 | 0 |
| FB | -13 | -9 | -14 | -12 | -23 | -28 | -36 | -32 | -35 | -30 | -28 | -20 |
| FBP | -18 | -24 | -2 | 6 | -7 | -12 | -14 | -1 | -58 | -32 | -17 | -16 |
| FBW | -110 | -63 | -49 | -11 | -45 | -36 | -29 | -2 | -153 | -120 | -89 | -61 |
| DS | 38 | 1 | 2 | 1 | 59 | 11 | 3 | 0 | 55 | -2 | -8 | -11 |
| SK | 35 | -3 | 6 | -4 | 33 | 26 | 10 | 8 | -3 | 20 | -5 | 0 |
| FL | -16 | -30 | 193 | 367 | 14 | 0 | 198 | 378 | -34 | -60 | 190 | 385 |
| LJ | 42 | 68 | 51 | 68 | 96 | 88 | 70 | 132 | -80 | 21 | -17 | 64 |
| OR | -78 | -43 | 36 | 29 | 28 | -4 | 34 | 37 | -128 | -94 | -41 | -38 |

Figure 7 presents the average update time. First, we can observe that the average update time of three workloads increases with the increase of $k$ as a larger $k$ incurs more subgraph computation costs. However, For DS and SK, the average deletion time drops when $k$ is 4 and 5. To explain, when $k$ is 4 and 5, the size of candidate cliques of DS and SK significantly decreases, which leads to the decreased cost of maintaining the candidate cliques. In addition, compared with building from scratch, our proposed insertion and deletion algorithms are efficient. For example, for OR, when $k$ is 3, the time of building from scratch equals that of 3.6M edge deletion operations, that of 48.6M edge deletion operations, and that of 6.5M mixed update operations. It shows that only with such a large amount of updates, the total update time is the same as that of building from scratch, which indicates the efficiency of our dynamic algorithms.

Besides, Table VIII shows the size of $\mathcal{S}$ after updates compared with that of building from scratch. We can observe that $\mathcal{S}$ preserves a high quality. For example, for OR after deletion when $k$ is 3, $\mathcal{S}$ decreases by 78, which is only 0.08% of the size of $\mathcal{S}$. For datasets like LJ, the size of $\mathcal{S}$ even increases. To explain, our swap operation can reach a nearly local optimum, and thus, its size may be larger than that of building from scratch. Thus, we can conclude that our proposed dynamic methods are efficient and effective.

## VII. CONCLUSIONS

In this paper, we study a new problem, *the maximum set of disjoint $k$-cliques*, which finds essential real-world applications in social networks but is proved to be NP-hard. To address that, we propose an efficient lightweight method that achieves a $k$-approximation to the optimal. Specifically, we develop several optimization techniques, including the node ordering method, the $k$-clique ordering method, and a lightweight implementation. Besides, to handle dynamic graphs, we devise an efficient indexing method with careful swapping operations, leading to the efficient maintenance of a near-optimal result with frequent updates in the graph. In various experiments on several large graphs, our proposed approaches significantly outperform the competitors in terms of both efficiency and effectiveness.